\documentclass[11pt]{article}

\usepackage[final]{acl}

\usepackage{times}
\usepackage{latexsym}
\usepackage{amsmath} 
\usepackage{amssymb} 
\usepackage{bbding}

\usepackage[T1]{fontenc}

\usepackage[utf8]{inputenc}

\usepackage{microtype}

\usepackage{inconsolata}

\usepackage{graphicx}
\usepackage[edges]{forest}
\usepackage{tikz}
\usetikzlibrary{positioning,shapes,arrows.meta}
\usepackage{array}
\usepackage{multirow}
\usepackage{booktabs}
\usepackage[dvipsnames]{xcolor}
\usepackage{comment}
\usepackage{enumitem}

\forestset{
  titlebox/.style={
    draw, rounded corners=2pt, fill=gray!35,
    font=\bfseries, align=center, minimum width=14mm
  },
  category/.style={
    draw, rounded corners=2pt, fill=gray!12,
    font=\bfseries, text width=40mm, align=left
  },
  subcat/.style={
    draw, rounded corners=2pt, fill=white,
    text width=40mm, align=left
  },
  refnode/.style={
    draw=blue!60!black, fill=blue!2, text=blue!60!black,
    font=\footnotesize\itshape,
    text width=65mm, align=left
  }
}
\definecolor{antiquefuchsia}{rgb}{0.57, 0.36, 0.51}

\newcommand{\MyListBold}[1]{%
  \par\smallskip\noindent\textbf{#1}\ %
}

\newcommand{\GreenCheck}{\textcolor{ForestGreen}{\CheckmarkBold}}
\newcommand{\RedX}{\textcolor{BrickRed}{\XSolidBrush}}

\title{Reward Engineering for Software Tasks: A Survey of Reinforcement Learning Approaches}

\author{
    \textbf{Md Rayhanul Masud}$^1$, 
    \textbf{Azmine Toushik Wasi}$^2$, 
    \textbf{Salman Rahman}$^3$, 
    \textbf{Md Rizwan Parvez}$^4$ \\
    $^1$University of California, Riverside \\
    $^2$Shahjalal University of Science and Technology \\
    $^3$University of California, Los Angeles \\
    $^4$Qatar Computing Research Institute (QCRI) \\
}

\begin{document}
\maketitle

\begin{abstract}
Reinforcement learning is increasingly used for code-centric software engineering tasks, including code generation, understanding, repair, testing, and optimization, especially with the rise of large language models and autonomous agents. 
A core challenge in these settings is reward design. 
Unlike standard RL domains with clear scalar objectives, software tasks involve competing goals such as correctness, security, efficiency, and readability, which are difficult to capture with a single reward. 
As a result, RL-for-SE systems rely on heterogeneous signals, including compilation results, unit tests, coverage metrics, retrieval scores, and learned preferences. 
Yet this work remains scattered across tasks and communities. 
This survey provides the first systematic review of reward engineering for RL in software tasks. We organize prior work by reward source, granularity, and aggregation. 
We then distill the findings into practical guidance, including a decision matrix, a design guide, and a reward engineering reporting standard for future RL-for-SE systems.
\end{abstract}

\section{Introduction}

Reinforcement learning has become an effective paradigm for software engineering tasks by enabling agents to learn from executable feedback, test outcomes, and preference signals rather than from labeled examples alone~\cite{wang2025survey}.
Recent systems, including CodeRL for program synthesis~\cite{le2022coderl}, DeepSeek-R1 for reasoning~\cite{guo2025deepseek}, and RL-based agents for automated program repair~\cite{wei2025swe}, have demonstrated state-of-the-art performance on code generation, bug fixing, test synthesis, and repository-level issue resolution across benchmarks such as HumanEval, MBPP, and SWE-bench.
A key factor underlying this progress is the design of the RL environment and reward function.
The environment specifies how the agent interacts with software artifacts through states, actions, and transitions~\cite{sutton1998reinforcement}, whereas the reward function maps complex software objectives, including correctness, efficiency, security, and readability, into scalar learning signals~\cite{wei2025swe}.

Despite the growing success of RL in software engineering, reward design remains a central challenge~\cite{wang2025survey}.
Unlike game-playing domains with clear, immediate rewards~\cite{mnih2015dqn,silver2018alphagozero}, software tasks involve competing criteria that resist reduction to a single scalar signal.
A program may compile yet fail tests, satisfy functional requirements yet expose security flaws, or behave correctly while remaining hard to understand or inefficient.
Existing approaches typically optimize only part of this space, relying on execution-based rewards for correctness, similarity-based rewards for reference alignment, or learned preference models for more subjective qualities.
When such signals are combined, additional challenges arise, including reward scaling, weight sensitivity, and conflicts among objectives~\cite{roijers2013survey,agarwal2022morl,deLangis2024dynamic,lu2025dynamic}.
These challenges are further shaped by the surrounding RL environment, since reward behavior depends on how states are represented, which actions are permitted, and at what temporal granularity feedback is assigned.

Given these challenges, a systematic and reward-centric survey of RL for software engineering is both timely and necessary, yet no such resource exists.
Existing surveys on RL for SE~\cite{yang2022survey, watson2022systematic, sharma2024survey, wang2025survey, zhang2025survey} primarily address algorithm selection, task performance, and application domains, treating reward functions as secondary implementation details rather than as first-class design choices that fundamentally shape learning outcomes.
Surveys on RLHF and RL for LLMs~\cite{ibrahim2024comprehensive, zhang2025survey} discuss reward modeling for alignment and reasoning, but do not address the unique challenges of software artifacts, namely executability, test coverage, semantic correctness, and deterministic oracle availability, which give rise to a qualitatively different reward design landscape.

To address this gap, we present the first systematic survey of reward design approaches to reinforcement learning in software engineering, covering 50+ papers published in 2018 - 2025.
We organize the literature through a taxonomy grounded in three reward design questions: what signal is used, at what temporal granularity it is delivered, and how multiple signals are aggregated.
We address the following research questions:

\MyListBold{RQ1:} How can reward schemes in RL for software engineering be categorized across tasks and mechanisms?

\MyListBold{RQ2:} What reward design patterns recur across different software engineering tasks?

\MyListBold{RQ3:} How are rewards defined at different granularities, from token and line-level to program and trajectory-level?

\MyListBold{RQ4:} How are heterogeneous reward signals, such as execution correctness, semantic similarity, and human preference, combined during training?

Unlike prior surveys, this work focuses on reward engineering as the core design issue. Because software supports executable, deterministic feedback, reward design can rely on tests, coverage, and execution traces.
A unifying principle emerges from this analysis: reward source, granularity, and aggregation choices are not free design parameters but follow from task structural conditions, namely oracle availability, observability of intermediate states, and the pressure to combine heterogeneous signals. 
We distill this principle into the decision matrix shown in Table~\ref{tab:taxonomy_compact} and revisit it across coding tasks in Section~\ref{sec:all-code-tasks}.
The survey organizes these reward choices, highlights common patterns and failures, and serves as both a reference and a practical design guide for RL in software engineering.
\label{sec:intro}

\section{Background}

\begin{table*}[!t]
\centering
\footnotesize
\renewcommand{\arraystretch}{1.28}
\caption{Taxonomy and decision matrix for surveyed RL-for-SE tasks, summarizing benchmarks, task conditions, primary rewards, aggregation strategies, and representative works. Per-paper details appear in Appendix~\ref{app:per-paper}.}
\label{tab:taxonomy_compact}
\begin{tabular}{@{}p{1.7cm} | p{2.8cm} | p{1.9cm} |p{2.5cm} |p{2.2cm} |p{2.8cm}@{}}
\toprule
\textbf{Task family} & \textbf{Benchmarks} & \textbf{Task condition} & \textbf{Primary reward (+ shaping)} & \textbf{Recommended aggregation} & \textbf{Representative works} \\
\midrule
\hline

NL2Code &
HumanEval(+), APPS, MBPP, EvalPlus, LiveCodeBench &
Executable, cheap oracle &
Execution; similarity / process-level shaping &
Constrained; correctness filter &
CodeRL, PPO-Coder, RLEF, PRL-Coder, Posterior-GRPO \\
\hline

Code Completion &
RepoBench, CrossCodeEval &
Partial or non-runnable target &
Similarity, retrieval-based proxy &
Normalized scalarization &
RLCoder, IRCoCo, aiXcoder-7B-v2 \\
\hline

Code2Code &
CodeXGLUE &
Executable &
Execution &
Constrained; correctness filter &
CoTran, EffiReasonTrans \\
\hline

Code2NL &
CodeSearchNet, TL-CodeSum &
Reference-based evaluation &
Similarity &
Normalized scalarization &
Hybrid2Seq+DRL, HAN, TAG \\
\hline

Code Retrieval &
CodeSearchNet, CoSQA &
Ranking-based success &
Retrieval; similarity as auxiliary &
Single signal preferred &
CoaCor, Cosoch, QueCos \\
\hline

Issue Resolution &
SWE-bench Lite, SWE-bench Verified &
Executable, expensive oracle &
Execution; gated by retrieval, similarity, or preference &
Cost-aware constrained &
SWE-Gym, DeepSWE, SWE-Swiss, R2E-Gym \\
\hline

Bug Localization &
Defects4J variants &
Ranking-based success &
Retrieval/ execution based coverage &
Single signal preferred &
RLocator, RecBi \\
\hline

Program Repair &
Defects4J, QuixBugs &
Executable &
Execution; preference shaping &
Constrained; correctness filter &
RePair, Repair-R1, Repairity \\
\hline

Vulnerability det.\,/\,repair &
Big-Vul, Devign, CVEFixes &
Limited or non-executable &
Similarity / retrieval + verifier &
Normalized scalarization &
ProRLearn, Vul-R2, SecRepair, RLFD \\
\hline

Test Gen.\,/\,Fuzzing &
HumanEval-Tests, V8, MBPP-Tests, Test262,  SpiderMonkey &
Executable, exploration-oriented &
Execution; coverage for fuzzing &
Constrained, with coverage shaping &
AceCoder, CURE, UTRL, CovRL-Fuzz, RLF, REINAM \\
\hline

\bottomrule
\end{tabular}
\end{table*}

\subsection{Reward Methods.}
\label{sec:reward-methods}
In RL for software engineering (SE), rewards map task goals to optimization signals, reflecting objectives such as correctness, coverage, security, readability, and human preference.
We distinguish three source families and five operational signal types.
The first family is \emph{verifiable tool-grounded rewards}, which derive from deterministic external checks such as compilation, execution outcomes, coverage, or other tool-based analyzers. 
In this family, \textbf{Execution} denotes correctness signals from compilation and tests, while \textbf{Coverage} denotes exploration-oriented signals such as line, branch, or mutation coverage.
The second family is \emph{reference-based proxy rewards}, which compare model outputs against gold artifacts or labeled targets. 
In this family, \textbf{Similarity} refers to lexical, structural, or semantic comparison metrics such as BLEU, CodeBLEU, AST similarity, or BERTScore, while \textbf{Retrieval} refers to ranking-oriented metrics such as MRR, MAP, NDCG, or Precision@k in retrieval and localization settings.
The third family is \emph{learned evaluators}, which provide reward signals through a learned scoring function rather than a deterministic oracle. 
In this family, \textbf{Preference} denotes human or model-aligned judgments over qualities such as usefulness, readability, or reasoning quality. 
Process reward models and learned critics used as reward proxies also belong to this family when they are used to provide optimization signals.

These reward families also vary along two cross-cutting axes that affect calibration, drift, and training stability~\cite{roijers2013survey}. 
First, reward sources differ in \emph{model dependence}: tool-grounded rewards are deterministic and model-independent, learned evaluators are model-dependent and may drift as models change~\cite{gao2023scaling}, while reference-based rewards fall between these extremes, since static metrics such as BLEU and CodeBLEU are model-independent whereas learned metrics such as BERTScore are model-dependent.
Second, they differ in \emph{adaptivity}: most works use static reward schedules, while adaptive designs condition one signal on another. Posterior-GRPO~\cite{fan2025posterior}, for example, applies preference rewards only after execution succeeds, so preference refines correct code rather than compensating for failures.

We use \textbf{Execution}, \textbf{Similarity}, \textbf{Retrieval}, \textbf{Coverage}, and \textbf{Preference} as shorthand labels throughout the paper.

\paragraph{RL Optimization Methods.}
Most works fine-tune policies with on-policy methods such as \emph{PPO}~\cite{schulman2017proximal}, whose clipped objective stabilizes updates under sparse, high-variance rewards. 
Recent alternatives pursue the same goal differently: \emph{GRPO}-style~\cite{shao2024deepseekmath} updates normalize rewards over sampled completion groups, reducing reliance on a value critic, while \emph{RLVR} (Reinforcement Learning with Verifiable Rewards)~\cite{lambert2024tulu} integrates deterministic feedback loops for heterogeneous, multi-objective optimization.

\paragraph{Coding Task Types Covered in This Survey.}
We discuss reward design across several SE tasks, including: code (i) \emph{generation}, (ii) \emph{understanding}, (iii) \emph{maintenance}, (iv) \emph{security}, and (v) \emph{testing}.

We summarize the surveyed corpus along task families, benchmarks, structural conditions, and predominant rewards in Table~\ref{tab:taxonomy_compact}, with per-paper reward-signal detailed in Appendix~\ref{app:per-paper} (Table~\ref{tab:code_tasks}).

\subsection{Survey Methodology.}
\label{sec:Survey-method}
\paragraph{Literature Collection.}
We conducted a broad literature search using Google Scholar, arXiv, and Semantic Scholar.
Because reward design for RL in SE spans both ML and software engineering, the search covered
major ML/RL venues such as NeurIPS, ICLR, ICML, and AAAI, as well as relevant SE and
programming-language venues. To capture fast-moving work, recent arXiv preprints were included,
and citation trails were followed through reference lists and \emph{cited by} links.

The search used keyword combinations around RL and software tasks, including
``reinforcement learning'', ``reward design'', ``code generation'', ``code maintenance'',
``vulnerability detection and repair'', and ``test generation''. The survey covers papers through
December 2025; given the pace of new preprints, some very recent manuscripts may have been missed.

\paragraph{Inclusion and Exclusion Criteria.}
We included papers that explicitly define, learn, or use reward signals for RL-driven
SE tasks, with enough detail to show how the reward guides optimization.
We excluded papers that apply RL to software problems but use only a generic task-success
signal without analyzing the reward design. Each candidate was screened through its
abstract, method, and experimental sections, focusing on reward formulation, granularity,
and multi-signal aggregation. The resulting paper set forms the basis for the taxonomy.

\label{sec:background}

\section{Code-Centric Tasks}
\label{sec:all-code-tasks}
We organize the survey around five categories of code-centric tasks: code generation, code understanding, code maintenance, code security, and code testing. As summarized in Table~\ref{tab:taxonomy_compact}, these tasks differ in their reward conditions, available benchmarks, primary reward signals, and aggregation strategies. The following subsections examine how these structural differences shape reward formulation within each category.

\subsection{Code Generation}


\subsubsection{NL2Code}
\label{sec:nl2code}

For NL2Code, the model maps natural-language specifications to executable programs, so correctness is ultimately defined by program behavior. 
As summarized in Table~\ref{tab:taxonomy_compact}, this makes NL2Code an executable task with relatively cheap oracle feedback, where \emph{execution} provides the most faithful reward whenever compiler or unit-test feedback is available.

CodeRL~\cite{le2022coderl}, B-Coder~\cite{yu2023mathcal}, and RLEF~\cite{gehring2024rlef} all follow this design, differing mainly in optimization: CodeRL uses graded penalties and a learned critic, B-Coder adopts a Dueling DQN formulation, and RLEF applies execution-based feedback to multi-turn refinement. Thus, their shared reward-design choice is functional correctness.

The main challenge is that execution feedback is often sparse. PPO-Coder~\cite{shojaee2023execution} mitigates this by adding AST and DFG-based \emph{similarity} to a reference solution as structural shaping. Other work introduces intermediate supervision while keeping final tests as the anchor: PRM-based RL~\cite{dai2024process} combines unit-test rewards with learned process rewards, and PRL-Coder~\cite{ye2025process} derives step-level feedback from compiler-verified mutation and refactoring. Posterior-GRPO~\cite{fan2025posterior} adds a \emph{preference} signal from a reasoning model, but conditions it on execution success so that reasoning is rewarded only when it leads to correct code.

\begin{center}
\fbox{
\parbox{0.93\linewidth}{
\textbf{Takeaway.}
Ground NL2Code reward design in \emph{execution} whenever an oracle is available. Use \emph{similarity} and process rewards only as shaping signals to reduce sparsity, while keeping them aligned with final test outcomes. Reserve \emph{preference} rewards for secondary goals such as reasoning quality, and condition them on execution correctness when possible.
}
}
\end{center}

\subsubsection{Code Completion}
\label{sec:codecompletion}

For code completion, the model generates continuations for partially observed programs, from local autocomplete to repository-level suggestions. Because these targets are often fragments rather than runnable programs, \emph{execution} is usually too sparse or costly as the main reward. Table~\ref{tab:taxonomy_compact} captures this partial-target condition, under which reward design relies instead on reference-based proxies, \emph{retrieval} signals, or learned \emph{preference} objectives.

RLCoder~\cite{wang2024rlcoder} follows this direction most directly, using a \emph{retrieval}-based proxy reward defined by how much retrieved context reduces the perplexity of the ground-truth completion. IRCoCo~\cite{li2024ircoco} addresses the same problem from a different angle, training a critic to regress reference-based metrics such as BLEU and Edit Similarity, thereby converting sparse sequence-level scores into token-level feedback. CoLT, the training framework behind aiXcoder-7B-v2~\cite{li2025aixcoder}, pushes this one step further by constructing synthetic preference pairs from reference code and optimizing them with DPO, replacing direct metric optimization with pairwise \emph{preference} learning.

\begin{center}
\fbox{
\parbox{0.93\linewidth}{
\textbf{Takeaway.}
For code completion, use reference-grounded proxy rewards when targets are partial or non-executable. Depending on the setting, these proxies may be \emph{retrieval}-based, similarity-based, or learned as \emph{preference} objectives from reference code.
}
}
\end{center}

\subsubsection{Code2Code}
\label{sec:code2code}

For code translation and migration, the goal is to convert programs across languages or APIs while preserving behavior. Since behavioral preservation can often be verified through compilation and testing, \emph{execution} is the natural primary reward. CoTran~\cite{jana2023cotran} uses compilation feedback for syntactic validity and test-based feedback for functional equivalence, while EffiReasonTrans~\cite{wang2025effireasontrans} likewise centers reward design on unit-test pass rate, adding only a length-based \emph{preference} term for efficiency. 

\begin{center}
\fbox{
\parbox{0.93\linewidth}{
\textbf{Takeaway.}
For code translation, use \emph{execution} as the primary reward whenever behavioral equivalence can be tested. \emph{Preference} terms can be added for secondary goals such as efficiency, but should remain guided by correctness.
}
}
\end{center}

\subsection{Code Understanding}

\subsubsection{Code2NL}

Code2NL generates natural-language descriptions of programs, such as summaries or docstrings. 
Because success is usually evaluated by agreement with reference text rather than by behavioral verification, RL methods in this setting typically rely on sequence-level \emph{similarity} rewards. Hybrid2Seq+Attn+DRL~\cite{wan2018improving} and HAN~\cite{wang2020reinforcement} both follow this design, assigning a terminal BLEU reward after the full description is generated. 
TAG~\cite{cai2020tag} extends the same idea through hierarchical RL, using BLEU-4 or ROUGE while combining the RL objective with MLE during training. 
Across these methods, the main limitation is that the rewards are sparse and tend to capture textual overlap more directly than semantic correctness.

\begin{center}
\fbox{
\parbox{0.93\linewidth}{
\textbf{Takeaway.}
For Code2NL, use \emph{similarity} as the primary reward, since success is defined by how closely the generated description matches the reference text.
}
}
\end{center}

\subsubsection{CodeRetrieval}

In code retrieval, the objective is to rank relevant code snippets for a given query, so reward design is most naturally tied to retrieval quality. CoaCor~\cite{yao2019coacor} makes this explicit: instead of optimizing for human-like annotations or BLEU overlap, it uses MRR from a retrieval model as the reward signal. Although similarity scoring appears inside that model, the optimized objective is clearly \emph{retrieval}-based. Cosoch~\cite{li2019reinforcement} follows the same direction by directly optimizing $\mathrm{NDCG}@10$ for session-level search. 
QueCos~\cite{wang2022enriching} adds BLEU-4 \emph{similarity} alongside MRR, but its reward design remains centered on ranking performance.
\emph{Similarity} helps as auxiliary shaping, but it is weaker than retrieval metrics as a measure of success.

\begin{center}
\fbox{
\parbox{0.93\linewidth}{
\textbf{Takeaway.}
For code retrieval, use \emph{retrieval} rewards such as MRR or NDCG as the primary objective. Add \emph{similarity} only as auxiliary shaping, since higher textual overlap does not necessarily improve ranking quality.
}
}
\end{center}

\subsection{Code Maintenance}

\subsubsection{Issue Resolution}
\label{sec:issueresolution}

Issue resolution is a long-horizon repository task in which the model must localize the bug, edit the relevant files, and produce a patch that satisfies repository-level constraints. 
Because success is ultimately determined by whether the patch works, \emph{execution} is the most faithful reward. 
SWE-Gym~\cite{pan2024training}, DeepSWE~\cite{luo2025deepswe}, and SWE-Swiss~\cite{SWESwiss2025} therefore stay centered on test-based correctness, with DeepSWE additionally using verifier-based \emph{preference} signals for test-time selection.

This corresponds to the issue-resolution setting in Table~\ref{tab:taxonomy_compact}: execution remains the most faithful reward, but repository-level testing cost motivates cheaper proxy signals for selection and stabilization.
SWE-RL~\cite{wei2025swe} replaces execution with a rule-based \emph{similarity} reward, 
while Satori-SWE~\cite{zeng2025satori} uses a staged design that combines \emph{retrieval} and learned \emph{preference} signals, with execution optionally used during selection.
R2E-Gym~\cite{jain2025r2e} occupies a middle ground, pairing execution-based testing with execution-free verifiers. 

\begin{center}
\fbox{
\parbox{0.93\linewidth}{
\textbf{Takeaway.}
For issue resolution, use \emph{execution} as the primary reward whenever repository tests are available. Add \emph{similarity}, \emph{retrieval}, or \emph{preference} only as auxiliary signals for scalability, ranking, or stability.
}
}
\end{center}

\subsubsection{Bug Localization}
\label{sec:bug-localization}

RL-based bug localization typically optimizes execution-derived or ranking-based signals rather than direct patch correctness. 
Table~\ref{tab:taxonomy_compact} captures this distinction: ranking-based localization aligns with \emph{retrieval} rewards, while test or trace-mediated localization relies on execution-derived coverage signals.
RecBi~\cite{chen2020enhanced} approaches compiler bug isolation through \emph{execution}-derived coverage shaping: it uses RL to guide the generation of passing test programs, then compares their execution traces with the failing test to localize buggy files. 
Here, coverage  acts as the shaping signal, while test outcomes provide the execution oracle. 
By contrast, RLocator~\cite{chakraborty2024rlocator} formulates bug localization directly as a ranking problem and optimizes \emph{retrieval} quality through metrics such as MRR and MAP.

\begin{center}
\fbox{
\parbox{0.93\linewidth}{
\textbf{Takeaway.}
For bug localization, prefer \emph{retrieval} rewards when the objective is ranking suspicious code elements, and opt for \emph{execution}-derived signals for settings where localization is mediated by test outcomes or coverage traces.
}
}
\end{center}

\subsubsection{Program Repair}
\label{sec:programrepair}

Program repair naturally favors \emph{execution}-based rewards because candidate patches can often be checked by compilation and tests. The main difficulty is sparse pass/fail feedback. RePair~\cite{zhao2024repair} addresses this with a critic trained on execution-derived signals, while Repair-R1~\cite{hu2025repair} adds rule-based rewards for test generation, repair, and output format. Repairity~\cite{tang2025boosting} takes a different route by using a learned \emph{preference} signal to rank candidate repairs, with execution retained mainly as a correctness filter during data collection.

\begin{center}
\fbox{
\parbox{0.93\linewidth}{
\textbf{Takeaway.}
For program repair, use \emph{execution} as the primary reward when patches are testable. Add critics, rule-based shaping, or \emph{preference} signals only to densify feedback or stabilize optimization.
}
}
\end{center}

\subsection{Code Security}

\subsubsection{Vulnerability Detection and Repair}
\label{sec:vulnerability}

Vulnerability detection and repair share a limited-oracle constraint: reliable \emph{execution} signals are often unavailable or weak. Detection methods therefore rely on label-based proxies. ProRLearn~\cite{Ren2024ProRLearnBP} uses reference-based \emph{similarity}, while RLFD~\cite{jiang2025enhancing} uses ranking rewards over vulnerable-line overlap and top-ranked accuracy, making it primarily \emph{retrieval}-based. In both cases, supervision comes from annotated labels rather than executable security outcomes, matching the limited-oracle regime in Table~\ref{tab:taxonomy_compact}.
On the repair side, PairCode~\cite{islam2024code} and SecRepair~\cite{islam2024llm} remains fully \emph{similarity}-based, combining CodeBLEU and BERTScore against the reference patch. 
Vul-R2~\cite{wen2025vul} and SecCoderX~\cite{wu2026secure} replace reference matching with learned verifier rewards for repair judgment and secure generation, respectively.


\begin{center}
\fbox{
\parbox{0.93\linewidth}{
\textbf{Takeaway.}
For vulnerability repair, use \emph{similarity} when execution signals are unavailable; add \emph{retrieval} for fine-grained localization, and use \emph{preference} signals when weak reference overlap makes similarity insufficient.
}
}
\end{center}
\subsection{Code Testing}

\subsubsection{Test Generation and Fuzzing}
\label{sec:testgenfuzzing}

Test generation and fuzzing both synthesize inputs to probe program behavior, but they optimize for different outcomes: test generation seeks valid, discriminative tests, while fuzzing emphasizes exploration and failure discovery. In both settings, rewards are naturally grounded in \emph{execution}. For test generation, CURE~\cite{wang2025co} uses execution outcomes from generated programs and tests to train a co-evolving test generator, and UTRL~\cite{lee2025learning} derives validity and discrimination rewards from pass/fail behavior. AceCoder~\cite{zeng2025acecoder} similarly supports binary execution rewards or learned \emph{preference} rewards from pass-rate comparisons.

Fuzzing follows the same execution-centered logic, but adds stronger exploration. RLF~\cite{su2022effectively} combines vulnerability discovery with \emph{coverage}, and CovRL-Fuzz~\cite{eom2024fuzzing} further rewards rarely covered regions while penalizing invalid inputs. 
Both sub-areas align with the executable-with-exploration setting in Table~\ref{tab:taxonomy_compact}: execution provides the correctness anchor, while coverage guides exploration.

\begin{center}
\fbox{
\parbox{0.93\linewidth}{
\textbf{Takeaway.}
For test generation and fuzzing, use \emph{execution} as the primary reward. Add \emph{coverage} when exploration matters, especially in fuzzing, and use learned \emph{preference} signals to refine execution feedback.
}
}
\end{center}

\label{sec:codetasks}

\begin{figure}[t]
    \centering
    \includegraphics[height=0.25\textwidth]{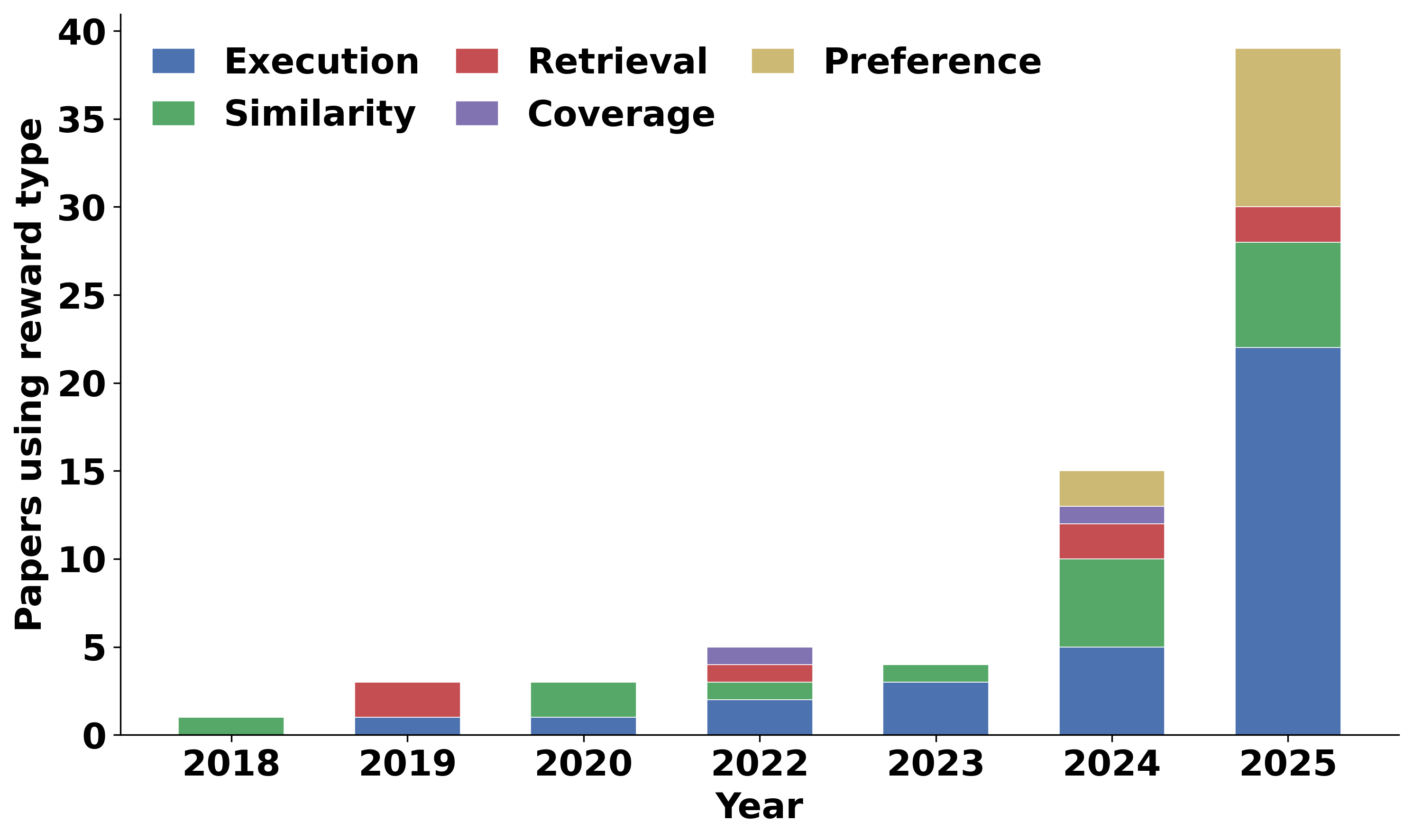}
    \caption{Number of papers per reward type by year.}
    \label{fig:temporal}
\end{figure}

\begin{table*}[t]
\centering
\footnotesize
\setlength{\tabcolsep}{4pt}
\caption{Aggregation strategies and adoption status in surveyed RL-for-SE work.}
\label{tab:aggregation}
\begin{tabular}{@{}p{0.23\linewidth} | p{0.13\linewidth} | p{0.55\linewidth}@{}}
\toprule
\textbf{Strategy} & \textbf{Status} & \textbf{Surveyed example} \\
\midrule
Fixed scalarization
  & Common
  & \emph{PPO-Coder}:fixed weighted sum of execution, AST, and DFG rewards \\
  \hline
Normalized scalarization & Rarely reported explicitly & Keeps execution, similarity, and preference scores comparable before combining them \\
  \hline
Pareto-aware aggregation
  & Absent
  & Remains an open direction for competing SE objectives such as correctness, efficiency, and security  \\
  \hline
Adaptive weighting
  & Nascent
  & \emph{Posterior-GRPO} conditions the preference reward on execution success \\
  \hline
Constrained aggregation
  & Implicit
  & \emph{AceCoder}: test outcomes constrain filtering and preference-pair construction \\
\bottomrule
\end{tabular}
\end{table*}

\section{Discussion}

\subsection{From taxonomy to decision guide}
\label{sec:synthesis}

This section turns the surveyed taxonomy into a four-step reward-design guide. 
Figure~\ref{fig:temporal} shows that the design space has grown sharply since 2022, with execution-based rewards remaining dominant and preference-based rewards expanding fastest.


\paragraph{(D1) Select the reward source by oracle availability.}
Reward source selection should start from the strongest oracle the task provides, since fidelity depends on what can be verified reliably and at what cost. 
When compiler, test, or execution feedback is cheap, use \emph{execution} as the reward anchor, as in CodeRL~\cite{le2022coderl}, RLEF~\cite{gehring2024rlef}, and Repair-R1~\cite{hu2025repair}. 
When execution is costly, as in long-horizon issue resolution with \emph{Satori-SWE}~\cite{zeng2025satori}, use it selectively as a gate, while \emph{retrieval} or \emph{preference} signals guide selection. 
When execution is unavailable or weak, choose the closest aligned proxy: \emph{retrieval} for ranking and localization, as in RLocator~\cite{chakraborty2024rlocator}; \emph{similarity} for reference-anchored outputs, as in SecRepair~\cite{islam2024llm}; and \emph{preference} for subjective judgments, as in Posterior-GRPO~\cite{fan2025posterior}. 
In proxy-based settings, reward fit should be checked directly: verify that the proxy tracks downstream success, rather than treating it as reliable by default.

\paragraph{(D2) Use hybrid rewards as a correctness-preserving decision guide.}
Hybrid rewards are useful because execution is faithful but often sparse, delayed, or costly, whereas proxy rewards are denser but may not preserve behavior. 
The guiding principle is to start with the most faithful available signal and add auxiliaries when needed. 
When execution is available but sparse, keep it primary and use similarity as shaping, as in \emph{PPO-Coder}~\cite{shojaee2023execution}. 
When tasks involve coupled failure modes, give each signal a distinct role: execution for validity, retrieval for file or context selection, and preference for decision quality, as in \emph{Satori-SWE}. 
When execution is unavailable, proxies such as similarity can be used, as in \emph{PairCode}~\cite{islam2024code}, but they should not be treated as correctness evidence. 
Finally, when multiple signals are combined, avoid weighted sums that let proxy scores offset failed execution; instead, gate, calibrate, or filter proxies with execution or verifier feedback, as in \emph{Posterior-GRPO}, and \emph{AceCoder}~\cite{zeng2025acecoder}. 
Table~\ref{tab:aggregation} summarizes these regimes and their adoption.

\begin{table*}[t]
\centering
\footnotesize
\renewcommand{\arraystretch}{1.}
\setlength{\tabcolsep}{4pt}
\caption{Recurring failure modes in RL-for-SE reward design.}
\label{tab:failure_modes}
\begin{tabular}{@{}p{0.18\linewidth} |p{0.2\linewidth}| p{0.52\linewidth}@{}}
\toprule
\textbf{Failure mode} & \textbf{Triggered by} & \textbf{Surveyed example} \\
\midrule
Proxy misalignment
  & Similarity rewards in place of execution
  & \emph{PairCode}, \emph{SecRepair}: rely on CodeBLEU and BERTScore; textual overlap need not reflect correctness \\
  \hline
Sparsity and oracle cost
  & Execution rewards over long rollouts
  & \emph{RLEF}, \emph{SWE-Gym}, \emph{DeepSWE}: sparse, high-variance test feedback; densify with process-level shaping or learned verifier \\
  \hline
Reward hacking and calibration drift & Learned preference rewards & 
\emph{Repairity}, \emph{aiXcoder-7B-v2}: use AI-feedback or DPO over generated pairs; need execution/downstream checks for reliability \\
\bottomrule
\end{tabular}
\end{table*}

\paragraph{(D3) Set granularity by reliable observability.}
Reward granularity should match the finest level at which feedback is trustworthy. 
If correctness is only checkable after a full artifact is produced, use program-level rewards, as in RLEF~\cite{gehring2024rlef} and Repair-R1~\cite{hu2025repair}. 
If intermediate states are reliably verifiable, use step or line-level rewards to improve credit assignment, as in PRL-Coder~\cite{ye2025process} and PRM-based scoring~\cite{dai2024process}. 
If success is defined over rankings, sets, or multi-step decisions, use trajectory-level rewards, as in RLocator~\cite{chakraborty2024rlocator} and RLFD~\cite{jiang2025enhancing}. 
If terminal feedback is faithful but sparse, use a hybrid design that keeps execution as the anchor and adds shaping or critic signals, as in \emph{PPO-Coder}~\cite{shojaee2023execution} and \emph{CodeRL}~\cite{le2022coderl}.

\paragraph{(D4) Align evaluation methodology with the reward family.}
Each reward family shapes validation choices, including benchmarks, optimizers, and evaluation metrics:
\begin{itemize}[leftmargin=*, itemindent=0pt, noitemsep, topsep=1pt]

\item \textbf{Execution}: executable-oracle benchmarks such as HumanEval, SWE-bench, and Defects4J; PPO with KL regularization or learned critics; pass@$k$ or test-success rates~\cite{le2022coderl,gehring2024rlef}.

\item \textbf{Similarity}: reference-anchored benchmarks such as CodeSearchNet and Big-Vul; BLEU, ROUGE, or BERTScore, often combined with MLE for stability ~\cite{cai2020tag}.

\item \textbf{Retrieval}: ranking benchmarks such as CodeSearchNet and CoSQA; MRR, MAP, or NDCG as both reward and evaluation metric~\cite{yao2019coacor,chakraborty2024rlocator}.

\item \textbf{Preference}: completion benchmarks such as RepoBench and CrossCodeEval; DPO or GRPO-style optimization; downstream task accuracy used to check reward-model reliability~\cite{li2025aixcoder}.
\end{itemize}

\subsection{Challenges and future directions}
\label{sec:challenges}

\paragraph{(C1) Recurring failure modes.}
The surveyed corpus reveals three connected failure modes in reward design, summarized in Table~\ref{tab:failure_modes}. 
First, dense proxy rewards improve learnability but can drift from behavior: \emph{PairCode} and \emph{SecRepair}~\cite{islam2024code,islam2024llm} use CodeBLEU/BERTScore-style rewards, yet textual overlap need not imply correct or secure code. 
Second, execution rewards are behaviorally faithful but can be sparse and costly over long rollouts, as in \emph{RLEF}~\cite{gehring2024rlef}, \emph{SWE-Gym}~\cite{pan2024training}, and \emph{DeepSWE}~\cite{luo2025deepswe}. 
Third, learned preference rewards offer flexible feedback but introduce risks of reward hacking and calibration drift, as in \emph{Repairity}~\cite{tang2025boosting} and \emph{aiXcoder-7B-v2}~\cite{li2025aixcoder}. 
A key future direction is therefore to make dense rewards more verifiable by grounding them in deterministic signals such as tests, static analyzers, or intermediate execution states.

\paragraph{(C2) Aggregation strategies remain under-explored.}
Table~\ref{tab:aggregation} shows that aggregation is still treated narrowly in surveyed RL-for-SE work. 
Fixed scalarization is common; \emph{PPO-Coder}~\cite{shojaee2023execution}, for example, combines execution, AST, and DFG similarity as a fixed sum. 
This design is simple but fragile: when signals differ in scale, noise, and reliability, a dense proxy can dominate sparse correctness feedback. 
More principled alternatives remain limited, with partial examples such as \emph{AceCoder}~\cite{zeng2025acecoder} for constrained aggregation and \emph{Posterior-GRPO}~\cite{fan2025posterior} for adaptive weighting.
By contrast, normalized scalarization is rarely reported explicitly, and Pareto-aware aggregation remains absent despite its relevance to multi-objective RL~\cite{roijers2013survey}. 
This gap matters because SE tasks often balance competing goals.
Future work should make normalization, weight sensitivity, and constraint handling explicit rather than hidden implementation choices.

\paragraph{(C3) Toward a Reward Engineering Reporting Standard.}
Existing works often report downstream performance without enough evidence about reward quality, making it difficult to separate genuine task progress from exploitation of a particular reward formulation. 
We therefore suggest a \emph{Reward Engineering Reporting Standard} with two linked parts. 
(a) \emph{Reward-weighting checklist}: authors should identify the task-faithful anchor reward, normalize heterogeneous components before aggregation, treat correctness-sensitive objectives as constraints so auxiliary rewards cannot compensate for failed checks, and report sensitivity to reward weights. 
(b) \emph{Reward-quality diagnostics}: when proxy and oracle signals are both available, authors should report their rank correlation, monitor reward--accuracy decoupling during training, and ablate each reward component against the downstream metric rather than only against the aggregate reward. 
Together, these practices make reward engineering more auditable and cross-system comparisons more reliable.

\label{sec:discussion}


\section{Conclusion}


This survey treats reward engineering for RL in software engineering as a first-class design problem. 
We organize prior work along three axes, reward source, granularity, and aggregation, and distill surveyed practice into a four-step design guide. 
A recurring pattern is that execution-grounded rewards provide the most dependable supervision, while similarity, retrieval, and preference rewards are most useful as shaping signals under sparse feedback rather than substitutes for correctness. 
This pattern also explains common failures, including proxy misalignment, reward sparsity, and unstable aggregation, and motivates principled aggregation, drift-resistant learned evaluators, and explicit reward-design reporting. 
Together, these directions can make RL-for-SE systems more reliable, comparable, and reproducible.
\label{sec:concl}

\section*{Limitations}

While this survey aims to provide a comprehensive view of reward design in reinforcement learning for software engineering tasks, the area is evolving rapidly and some very recent or adjacent works may have been missed. We intentionally scope our review to \emph{reward schemes used to train or adapt policies} for SE tasks (e.g., code generation, repair, translation, retrieval/repo-level completion, testing, and localization), and do not attempt to cover reward design for large-scale pre-training or general instruction tuning. 
We prioritize works where reward formulation is a central contribution. Works that apply RL only as a minor component, or where the reward is implicit/standard and not analyzed, are omitted. To help keep pace with new literature, we plan to maintain an accompanying repository that can be incrementally updated as the field progresses.
Lastly, this paper is a survey or review of existing research on reward design for reinforcement learning in software engineering tasks. Accordingly, it does not introduce new models or experimental evaluations; instead, it focuses on organizing, summarizing, and critically analyzing prior work in this area.
\label{sec:limitations}


\bibliography{latex/008-bib/papers}

\clearpage

\appendix
\section{Per-Paper Reward-Signal Map}
\label{app:per-paper}

\begin{table*}[!t]
\centering
\scriptsize
\caption{Overview of RL in SE tasks and their reward types. \GreenCheck\ indicates the paper uses the reward type; \RedX\ indicates it does not.}
\label{tab:code_tasks}
\renewcommand{\arraystretch}{1.3} 
\begin{tabular}{|p{1.4cm} | p{1.8cm} | c | c | c | c | c | c | c |}
\toprule
\textbf{} & \textbf{Code Task} & \textbf{Paper} & \textbf{Execution} & \textbf{Similarity} & \textbf{Retrieval} & \textbf{Coverage} & \textbf{Preference}\\
\midrule
\multirow{12}{*}{Generation}   & NL2Code    & CodeRL~\cite{le2022coderl} & \GreenCheck & \RedX & \RedX & \RedX & \RedX\\
                                   &                    & PPO-Coder~\cite{shojaee2023execution} & \GreenCheck & \GreenCheck & \RedX & \RedX & \RedX\\
                                   &                    & B-Coder~\cite{yu2023mathcal} & \GreenCheck & \RedX & \RedX & \RedX & \RedX\\
                                   &                    & RLEF~\cite{gehring2024rlef} & \GreenCheck & \RedX & \RedX & \RedX & \RedX\\
                                   &                    & PRM-Based~\cite{dai2024process} & \GreenCheck & \RedX & \RedX & \RedX & \RedX\\
                                   &                    & PRLCoder~\cite{ye2025process} & \GreenCheck & \RedX & \RedX & \RedX & \RedX\\
                                   &                    & Posterior-GRPO~\cite{fan2025posterior} & \GreenCheck & \RedX & \RedX & \RedX & \GreenCheck\\
                                   
                                   \cline{2-8}
                                   & Code Completion    & IRCoCo~\cite{li2024ircoco} & \RedX & \GreenCheck & \RedX & \RedX & \RedX\\
                                   &                    & RLCoder~\cite{wang2024rlcoder} & \RedX & \RedX & \GreenCheck & \RedX & \RedX\\
                                   &                    & aiXcoder-7B-v2~\cite{li2025aixcoder} & \RedX & \RedX & \RedX & \RedX & \GreenCheck\\
                                   \cline{2-8}
                                   & Code2Code   & CoTran~\cite{jana2023cotran} & \GreenCheck & \RedX & \RedX & \RedX & \RedX\\
                                   &                    & EffiReasonTran~\cite{wang2025effireasontrans} & \GreenCheck & \RedX & \RedX & \RedX & \GreenCheck\\
                                   
\midrule
\multirow{6}{*}{Understanding}  & Code2NL         & Hybrid2Seq+Attn+DRL~\cite{wan2018improving} & \RedX & \GreenCheck & \RedX & \RedX & \RedX\\
                               &                    & HAN~\cite{wang2020reinforcement} & \RedX & \GreenCheck & \RedX & \RedX & \RedX\\
                               &    & TAG~\cite{cai2020tag} & \RedX & \GreenCheck & \RedX & \RedX & \RedX\\
                               \cline{2-8}
                               & Code Retrieval   & CoaCor ~\cite{yao2019coacor} & \RedX & \RedX & \GreenCheck & \RedX & \RedX\\
                               &                    & Cosoch ~\cite{li2019reinforcement} & \RedX & \RedX & \GreenCheck & \RedX & \RedX\\
                               &                    & QueCos ~\cite{wang2022enriching} & \RedX & \GreenCheck & \GreenCheck & \RedX & \RedX\\
\midrule 
\multirow{14}{*}{Maintenance} & Issue Resolution         & SWE-RL~\cite{wei2025swe} & \RedX & \GreenCheck & \RedX & \RedX & \RedX\\
                               &                    & SWE-GYM~\cite{luo2025deepswe} & \GreenCheck & \RedX & \RedX & \RedX & \RedX\\
                               &                    & DeepSWE~\cite{luo2025deepswe} & \GreenCheck & \RedX & \RedX & \RedX & \GreenCheck\\
                               &                    & SWE-Swiss~\cite{SWESwiss2025} & \GreenCheck & \RedX & \RedX & \RedX & \RedX\\
                               &                    & Satori-SWE~\cite{zeng2025satori} & \GreenCheck & \RedX & \GreenCheck & \RedX & \GreenCheck\\
                               &                    & R2E-Gym~\cite{jain2025r2e} & \GreenCheck & \RedX & \RedX & \RedX & \RedX\\
                               &                    & Self-play SWE-RL~\cite{wei2025toward} & \GreenCheck & \RedX & \RedX & \RedX & \RedX\\
                               &                    & SWE-PRM~\cite{gandhi2025agents} & \GreenCheck & \RedX & \RedX & \RedX & \GreenCheck\\
                               \cline{2-8}
                               & Code Refactoring   & Refactor-RL~\cite{palit2024generating} & \GreenCheck & \RedX & \RedX & \RedX & \RedX\\
                               \cline{2-8}
                               &                    & CodeMentor~\cite{nashaat2024towards}  & \RedX & \RedX & \RedX & \RedX & \GreenCheck\\
                               & Code Review   & CRScore++~\cite{kapadnis2025crscore++} & \RedX & \RedX & \RedX & \RedX & \GreenCheck\\
                               \cline{2-8}
                               & Code Optimization   & Pearl~\cite{lamouri2025pearl} & \GreenCheck & \RedX & \RedX & \RedX & \RedX\\
                               \cline{2-8}
                               & Bug Localization   & RecBi~\cite{chen2020enhanced} & \GreenCheck & \RedX & \RedX & \RedX & \RedX\\
                               &                    & RLocator~\cite{chakraborty2024rlocator}  & \RedX & \RedX & \GreenCheck & \RedX & \RedX\\
                               \cline{2-8}
                               & Program Repair   & RePair~\cite{zhao2024repair} & \GreenCheck & \RedX & \RedX & \RedX & \RedX\\
                               &                    & RePair-R1~\cite{hu2025repair} & \GreenCheck & \RedX & \RedX & \RedX & \RedX\\
                               &                    & Repairity~\cite{tang2025boosting} & \GreenCheck & \RedX & \RedX & \RedX & \GreenCheck\\
\midrule
\multirow{5}{*}{Security} & Vulnerability & ProRLearn~\cite{Ren2024ProRLearnBP} & \RedX & \GreenCheck & \RedX & \RedX & \RedX\\
                               &      Detection              & RLFD~\cite{jiang2025enhancing} & \RedX & \RedX & \GreenCheck & \RedX & \RedX\\
                               \cline{2-8}
                               \cline{2-8}
                               & Vulnerability        & PairCode~\cite{islam2024code}  & \RedX & \GreenCheck & \RedX & \RedX & \RedX\\
                               &        Repair            & SecRepair~\cite{islam2024llm}  & \RedX & \GreenCheck & \RedX & \RedX & \RedX\\
                               &                    & Vul-R2~\cite{wen2025vul}  & \RedX & \GreenCheck & \RedX & \RedX & \GreenCheck\\
                               & 
                               & SecCoderX~\cite{wu2026secure} & \RedX & \GreenCheck & \RedX & \RedX & \GreenCheck\\
\midrule
\multirow{6}{*}{Testing} & Test Generation & AceCoder~\cite{zeng2025acecoder} & \GreenCheck & \RedX & \RedX & \RedX & \GreenCheck\\
                               &                    & CURE~\cite{wang2025co} & \GreenCheck & \RedX & \RedX & \RedX & \RedX\\
                               &                    & UTRL~\cite{lee2025learning} & \GreenCheck & \RedX & \RedX & \RedX & \RedX\\
                                \cline{2-8}
                               \cline{2-8}
                               & Fuzzing        & REINAM~\cite{wu2019reinam} & \GreenCheck & \RedX & \RedX & \RedX & \RedX\\
                               &                    & RLF~\cite{su2022effectively} & \GreenCheck & \RedX & \RedX & \GreenCheck & \RedX\\
                               &                    & CovRL-Fuzz~\cite{eom2024fuzzing} & \RedX & \RedX & \RedX & \GreenCheck & \RedX\\
\midrule 
\multirow{7}{*}{Misc}  &      Text2SQL              & CogniSQL-R1-Zero ~\cite{gajjar2025cognisql} & \GreenCheck & \RedX & \RedX & \RedX & \RedX\\
&                    & Think2SQL ~\cite{papicchio2025think2sql} & \GreenCheck & \RedX & \RedX & \RedX & \RedX\\
&                    & Reasoning-SQL ~\cite{pourreza2025reasoning} & \GreenCheck & \GreenCheck & \RedX & \RedX & \GreenCheck\\
                        \cline{2-8}
&      Visual2Code              & UI2Code ~\cite{yang2025ui2code} & \GreenCheck & \GreenCheck & \RedX & \RedX & \RedX\\
&                    & VinciCoder ~\cite{zhao2025vincicoder} & \GreenCheck & \GreenCheck & \RedX & \RedX & \RedX\\
&                    & ChartMaster ~\cite{tan2025chartmaster} & \GreenCheck & \GreenCheck & \RedX & \RedX & \RedX\\
        \cline{2-8}
&  & RL-GPT~\cite{liu2024rl} & \GreenCheck & \GreenCheck & \RedX & \RedX & \RedX\\
                               
\bottomrule
\end{tabular}
\end{table*}

Table~\ref{tab:code_tasks} reports reward-signal assignments for each surveyed paper, organized by code-task family. 
Each row marks the reward families used by a paper, while Table~\ref{tab:taxonomy_compact} summarizes the same assignments at the task-family level.

\section{Adjacent Task Families}
\label{app:misc}

Beyond the core code-centric tasks surveyed in Section~\ref{sec:all-code-tasks}, similar reward-design patterns appear in adjacent code-generation settings. 
In Text-to-SQL, CogniSQL-R1-Zero~\cite{gajjar2025cognisql}, Think2SQL~\cite{papicchio2025think2sql}, and Reasoning-SQL~\cite{pourreza2025reasoning} use \emph{execution} rewards for query correctness, with Reasoning-SQL further adding syntax, schema-linking, and LLM-judge signals as auxiliary shaping, paralleling NL2Code in Section~\ref{sec:nl2code}. 
In visual-to-code tasks, UI2Code~\cite{yang2025ui2code}, VinciCoder~\cite{zhao2025vincicoder}, and ChartMaster~\cite{tan2025chartmaster} combine visual \emph{similarity} rewards on rendered outputs with \emph{execution} penalties for non-compilable or non-renderable code, mirroring the execution-plus-similarity pattern in code translation discussed in Section~\ref{sec:code2code}. 
RL-GPT~\cite{liu2024rl} extends this pattern to embodied code generation through task-success \emph{execution} rewards with dense vision-language similarity shaping.

\section{Acknowledgements}
\paragraph{AI Assistance Disclosure.}
We used generative AI tools, including ChatGPT and Claude, for language editing and paraphrasing of author-written text. 
All content was reviewed and revised by the authors, who take full responsibility for the paper's correctness and citations.

\end{document}